# Inclusive Mentoring: The Mindset of an Effective Mentor

*Chandralekha Singh,*

*Department of Physics and Astronomy, University of Pittsburgh, Pittsburgh, PA 15260*

Mentoring is the process of forming, cultivating and maintaining relationship that supports and advances the mentees in their pursuits [1-2]. As physicists, we mentor undergraduate and graduate students in diverse settings, e.g., when we teach them in various courses, when we advise students in their research or when we advise them about academic and non-academic issues. For example, we give advice on what courses to take, whom to do research with, how to live a balanced life and manage academic and non-academic responsibilities, how to apply for financial supports, scholarships and jobs.

While effective mentoring can improve the outcome for all students in research and education, appropriate mentoring of students plays a critical role in ensuring that students from underrepresented groups thrive since physics is one of the least diverse STEM disciplines with stereotypes related to who can excel in it [3-11].

Research suggests that there is no single mentoring style that impacts effectiveness [2]. However, genuine concern for the mentee, boosting their sense-of-belonging and self-efficacy [4-6], inculcating growth mindset (intelligence in not immutable since your brain is like a muscle and can grow with hard work) [7,11], supporting students in learning to use effective strategies for growth that build on their current knowledge meaningfully, encouraging them to take advantage of their peers and mentors while helping them embrace their struggles as stepping stones to learning, are critical aspects of successful mentoring. It is important for mentors to convey to students that they have high expectations of them, they know that they have what it takes to excel if they work hard and work smart (which entails using effective approaches) and they are there to support them as needed. Observed outcomes of good mentoring are mentees more likely to succeed in their course work, research, be visible compared to others at a particular stage of their career, have a career plan in addition to being happy with their work because mentoring helps shape the outlook and attitudes of the mentee positively [1,2]. Mentoring can support growth in physics knowledge and skills as well as development of leadership skills while creating a climate of positivity, sense of competence and "you can do it" attitude [1,2].

Keeping in mind the role of mentors, a crucial characteristic of effective mentoring relationships in both teaching and advising contexts is that the mentors themselves have a "growth" mindset in order to help their mentees embrace struggles with physics and learn how to react to challenges and adversity. A research study involving the fixed vs. growth mindset of STEM instructors of undergraduate students was carried out at a large research university in the US in which 150 instructors from 13 different science disciplines including physics participated [11]. Instructors were asked to respond to the following two questions on a Likert scale (strongly agree, agree, disagree, strongly disagree etc.): "To be honest, students have a certain amount of intelligence, and they really can't do much to change it" and "Your intelligence is something about you that you can't change very much". Instructors who agree with these statements have a "fixed" mindset about their students' ability and those who disagree with these statements have a "growth" mindset about their students' ability. It was found that STEM instructors who had fixed mindset about students' ability in that they believed that students in their courses came with fixed abilities had larger achievement and motivation gaps. In particular, the achievement gaps between the racial and ethnic minority students and white students were twice as large and underrepresented students had worse motivational beliefs in the classes taught by instructors with a fixed mindset compared to instructors with a growth mindset.

It is not surprising that those in mentoring roles, e.g., instructors and research advisors, who have a fixed mindset about their students' ability are unlikely to inculcate growth mindset amongst their students, i.e., have students realize that intelligence is not fixed but can grow when one works hard and works smart and inspire them to use effective strategies to grow [7,10]. On the other hand, good mentors can shape the way that mentees react to challenges and ensure that the mentees embrace struggles while solving physics problems or conducting physics research as normal, unavoidable and an opportunity for developing expertise using deliberate strategies and build on their prior knowledge and skills effectively [7]. A good mentor can disambiguate student experiences in challenging situations and promote growth mindset and the importance of struggling in learning and excelling [7]. In particular, only when we as mentors have a growth mindset and truly believe that the students we are mentoring in instructional or research contexts can overcome challenges and excel by working hard and using effective strategies while we continue to support them, will students develop a mindset that challenges are not unique to them or permanent but universal and temporary [7]. As mentors, we should set high expectations for students but also provide assurances that students can reach those expectations by working hard and using effective approaches, struggling at the task and using their struggles as a learning opportunity while we continue to support them. Also, we should simultaneously help students learn effective cognitive and meta-cognitive strategies to scaffold their knowledge and skill development. What is important to recognize is that setting high standards *without* providing assurance that all students in a physics course or in a physics research lab can achieve them by working hard while using effective strategies can hurt those from the underrepresented groups (e.g., women and racial and ethnic minority students) the most. Due to the stereotypes associated with physics, without assurance from the mentor, those students are more likely to attribute their struggle to their lack of ability as opposed to a normal part of developing expertise [7].

Mentors should also positively recognize and praise students for making progress even if the mentor perceives them to be small steps, otherwise, lack of recognition is more likely to negatively impact those who are underrepresented in physics. Since physics is a field with strong stereotypes about who can succeed in it, these micro-affirmations are particularly important. Lack of positive recognition is known to negatively impacted the entry and retention of underrepresented students, e.g., women and racial and ethnic minority students in physics related disciplines for decades [3-7].

Eileen Pollock, the first woman to get a BS degree in physics at Yale University, decided to pursue graduate work in English and became an English professor at the University of Michigan despite finishing her physics undergraduate degree summa cum laude. In her memoir [12], she recounts the negative impact of lack of positive recognition from her thesis advisor, "Not even the math professor who supervised my senior thesis urged me to go on for a Ph.D. I had spent nine months missing parties, skipping dinners and losing sleep, trying to figure out why waves — of sound, of light, of anything — travel in a spherical shell, like the skin of a balloon, in any odd-dimensional space, but like a solid bowling ball in any space of even dimension. When at last I found the answer, I knocked triumphantly at my adviser's door. Yet I don't remember him praising me in any way. I was dying to ask if my ability to solve the problem meant that I was good enough to make it as a theoretical physicist. But I knew that if I needed to ask, I wasn't." She adds, "[I was] certain this meant I wasn't talented enough to succeed in physics, I left the rough draft of my senior thesis outside my adviser's door and slunk away in shame. Pained by the dream I had failed to achieve, I locked my textbooks, lab reports and problem sets in my father's army footlocker and turned my back on physics and math forever." This example illustrates a missed opportunity in which an undergraduate thesis mentor failed to positively recognize the accomplishment of a woman in physics and she went from feeling "triumphant" about having solved her thesis problem to feeling she wasn't talented enough to succeed in physics, otherwise her advisor would have praised her for her successful solving of the problem. What is also worth reflecting upon is that decades later while writing her book, when Pollock asked her thesis advisor what he thought of her thesis, he noted, "It's very unusual for any undergraduate to do an independent project in mathematics. By that measure, I would have to say that what you did was exceptional." She then asked if he ever specifically encouraged any undergraduates to go on for Ph.D.'s; after all, he was then the director of undergraduate studies. He said he never encouraged anyone. This type of lack of encouragement and positive recognition as someone who can excel in physics is likely to be particularly detrimental to the underrepresented students, e.g., women, because they do not have many peers and role models and are struggling with the idea of whether they belong in physics and have what it takes to excel in physics!

Due to the pervasive societal stereotypes associated with who belongs in physics and can excel in physics, many students from the underrepresented groups question whether they have what it takes to be a successful physicist [13-21]. Depending upon students' background and privilege, students may come to our classes and in our research labs with different motivational beliefs and prior preparations. Mentoring plays a critical role in ensuring that all students and especially those from the underrepresented groups receive adequate guidance, support, recognition and assurance to participate and excel in physics. Developing the mindset of an effective mentor requires us to be reflective and introspective and is crucial for ensuring that our students develop a growth mindset, embrace their struggles in physics and use effective approaches to overcome the struggles and excel in physics [10,22].